\documentclass{tMPH2e}

\usepackage{mathrsfs}
\usepackage{graphicx}
\usepackage{epstopdf}

\begin{document}

	\title{\bf Time-dependent local-to-normal mode transition in triatomic molecules}

	\author{Hans Cruz$^{a}$$^{\ast}$\thanks{$^\ast$Corresponding author. Email: hans@ciencias.unam.mx
	\vspace{6pt}}, 
	Marisol Berm\'udez-Monta\~na $^{a}$, 
	Renato Lemus$^{a}$\\
	\vspace{6pt}		
	$^{b}${\em{Instituto de Ciencias Nucleares, Universidad Nacional Aut\'onoma de M\'exico,\\ A. P. 70543, M\'exico, 		DF 04510, M\'exico.}}\\		
	\vspace{6pt}
	\received{v4.5 released September 2009} }

\maketitle

\begin{abstract}

Time-evolution of the vibrational states of two interacting harmonic oscillators in the local mode scheme is presented.  A local-to-normal mode transition (LNT) is identified and studied from the time dependent point of view. The LNT is established as a polyad breaking phenomenon from the local point of view for  the   stretching degrees of freedom in a triatomic molecule. 
This study is carried out  in the algebraic representation of bosonic operators. The dynamics of the states are determined via the solutions of the corresponding nonlinear Ermakov equation. Application of this formalism  to  H$_2$O, CO$_2$, O$_3$ and NO$_2$ molecules in the adiabatic, sudden and linear regime is considered.

\end{abstract}

\bigskip


\section{Introduction}

For a long time the normal mode picture (NM) to describe the molecular vibrational degrees of freedom was the unique fundamental tool to interpret  infrared and Raman spectra \cite{herzberg}. In this theoretical framework the system is described  as a set of   harmonic oscillators, whose non diagonal interactions 
give rise to the concept of normal polyad $\hat P_N$, a pseudo quantum number  encompassing subsets of interacting  states determined by the most relevant resonances \cite{kellmanP}. The normal mode analysis is conveniently treated in an algebraic manner through the introduction of bosonic operators \cite{califano,praga,kellman}.

 As the spectroscopic techniques become more refined \cite{yamanouchi1,yamanouchi2,yamanouchi3,nesbitt,field} highly excited states near the chemical "significant" region were detected, showing unexpected regularities (energy doublets)  from the point of view of the NM picture without resonances. This pattern however was well understood considering  a local mode model, which describes  the molecular system in terms of a    set of interacting local oscillators, typically associated  with the  stretching bonds involving large mass differences of type X-H, for instance \cite{hayward1,hayward2,hayward3,lawton,child,halonen,halonen2p,jensen2}. Abundant literature emerged concerned with this description as well as with the connection  with the traditional description in terms of normal modes \cite{mills,mills2,valle,lehmann,duncan1,duncan2,gambi,lehmann2,kellman2}.

The physical foundation behind the local mode description of a molecular system is the appearance of bonds involving large mass differences, which leads to the possibility of considering  the vibrational description  as a set of independent local oscillators at zeroth order, a natural description when both kinetic and potential energies are expanded in terms of local coordinates \cite{carrington}. This approach simplifies the vibrational description  allowing simple models to be applied. This contrasts to variational methods where the kinetic energy is calculated in exact form although computationally expensive.

 The suitability for applying a local  model may be pondered by the  splitting between the fundamentals associated with an equivalent set of vibrational degrees of freedom.  It is expected a small splitting relative to the fundamental energies to identify local behaviour. A remarkable feature of a local mode behaviour is that a local polyad $\hat P_{\tiny\mbox{L}}$ can be identified in terms local number operators besides the normal polyad $\hat P_{\tiny\mbox{N}}$. Due to the interaction between the oscillators  and  anharmonicity   a splitting among the levels associated with a given polyad is present. This splitting is a measure of the locality degree \cite{child,halonen}. This local polyad preserving feature characterizes the local mode behaviour.  As the interaction increases the levels associated with different polyads  approach and $\hat P_{\tiny\mbox{L}}$  stop being preserved. This interaction enhancement may be interpreted either as a geometry  and/or masses ratio changes. In this situation $\hat P_{\tiny\mbox{L}}$ is  broken and the normal mode description with well defined polyad $\hat P_{\tiny\mbox{N}}$ should be  preferred. This phenomenon is manifested as a transition process of polyad breaking but also by the  impossibility of a reasonable estimation of  the force constants starting from a local scheme \cite{lemusrev}.  This local-to-normal mode transition (LNT)
 has been studied taking the parametric transition from H$_2$O and CO$_2$ molecule \cite{marisol}.  The local polyad breaking has been detected using several concepts: probability density, fidelity, entropy and Poincar\'e sections.
This polyad breaking can be identified by considering two local interacting harmonic oscillators up to second order. In this case the local and normal polyads are related by an  expression of the form $\hat P_{\tiny\mbox{N}}=\zeta_0+\beta_0 \hat{P}_{\tiny\mbox{L}}+\gamma \hat{V}$, where
 $\zeta_0$ and $\gamma$ vanish in the local mode limit. The aim of this work is to study the LNT when $\hat{P}_{\tiny\mbox{L}}=\hat{P}_{\tiny\mbox{N}}$ stop being valid from the point of view of a time-dependent problem. 

 The use of exactly solvable models have pervaded  all  branches of physics. Time-independent Hamiltonian systems for instance present energy conservation, which implies that any one dimensional problem is  integrable. In a more general situation, systems with $n$-degrees of freedom that  possess $n$-constants of motion have separable time-independent Schr\"odinger equation~\cite{gutzwiller1990chaos}, a feature that allows us to solve the corresponding differential equation in coordinate representation or to construct the corresponding propagator. Examples where this situation applies are: free particle, harmonic oscillator and  motion of charged particles in uniform  electromagnetic fields, whose solutions were obtained since the early days of quantum mechanics~\cite{kennard1927quantenmechanik}.

In contrast, dealing with  time-dependent systems represents a more complicated situation even for  one dimensional systems since the Hamiltonian is no longer a constant of motion. In particular, because of its wide range of applications the study of the dynamics to the parametric oscillator,  i.e. an oscillator with time-dependent mass and frequency, has been the subject of major interest, leading to the search of  exact or adiabatic invariants. An adiabatic invariant for  the parametric oscillator problem was first introduced by {\it Lorentz} at the Solvay Congress on 1911~\cite{leach2008ermakov}. Later on, in $1966$ {\it Lewis} obtained in explicit form an exact class of  time-dependent invariants\footnote{Nowadays  in the literature this invariant is known as the {\it Ermakov invariant} in classical mechanics or also as the {\it Ermakov operator} in the field of  quantum mechanics, due to the fact that such invariant was reported by Ermakov in 1880 \cite{ermakov}. However, the Ermakov system was actually discussed six years earlier in 1874 by Steen~\cite{steen}. But since the latter work  was published in Danish in a journal not usually containing articles on mathematics, this contribution  
went unnoticed until the end of last century. An English translation of the original paper and generalizations can be found in Ref. \cite{redheffer1999steen, redheffer2001steen}.} for time-dependent one dimensional harmonic oscillator in both classical and quantum mechanical   schemes \cite{lewis}. {\it Lewis} itself together with {\it Riesenfeld } developed the theory of time-dependent invariants for non-stationary quantum systems,  applying the developed method to the one dimensional time dependent harmonic oscillator  as well as  to a charged particle  in a time-dependent electromagnetic field~ \cite{lewis1969exact}. 

On the other hand, two years after  Lewis' contribution,  linear invariant operators for time-dependent oscillator were introduced by Malkin, Man'ko and Trifonov~\cite{malkin1970coherent, malkin1973linear}. These operators can be used to define generalized correlated states, which is  highly convenient in many problems of quantum theory, with the additional advantage that  they can be found for any $N$-dimensional time-dependent quadratic system. 

The time evolution of quantum systems is still an important subject of research. A recent contribution proved the close relation between the dynamics  and the solutions of the Ermakov and Ricatti non-linear differential equations,  allowing a deeper physical insight into the systems~\cite{schuch2007connection,schuch2008riccati,Hans}. Additionally, the Ermakov invariant method for solving the quantum time-dependent parametric  oscillator has been connected with the so called quantum Arnold transformation, proving that both methods are equivalent~\cite{guerrero2013quantum, lopez2014generalizations, guerrero2015lewis}. In this contribution we shall analyze the LNT from a time-dependent point of view using the invariant operator theory. Particular attention is paid to the time-dependent polyad breaking transition by identifying the time dependent local number operators.

This article is organized as follows.  Section 2  is devoted to analyze  two interacting oscillators in order to establish the LNT from the local polyad breaking point of view. The aim of Section 3 is to introduce the  time dependent counterpart of the LNT.
In Section 4 the results of several molecular systems are presented for three cases: sudden, linear and adiabatic time-dependent evolution. Finally, in Section 5 the summary and our conclusions are presented.


\section{Stationary description of vibrational states}

\label{section-2}

In  general, a molecular  vibrational description may be carried out in local or normal mode schemes. The former case is appropriate for subspaces of equivalent oscillators where large mass ratios are present. Here we shall start considering two interacting local oscillators associated with the stretching degrees of freedom of triatomic molecules.

A  molecule suitable to be described in a local mode scheme presents a relatively small splitting among  the associated fundamentals. In this case a  good approximation consists in a zeroth order Hamiltonian $\hat{H}_0^{\tiny \mbox{loc}}$ corresponding to  two independent oscillators plus an interaction $\hat{V}_{\tiny \mbox{int}}^{\tiny \mbox{loc}}$ which is expected to yield a small contribution to the energy spectrum.
 Hence the Hamiltonian may be written in the form
	\begin{equation} \label{Hloc}
	\hat{H}_{\tiny\mbox{L}}=\hat{H}_0^{\tiny \mbox{loc}}+\hat{V}_{\tiny \mbox{int}}^{\tiny \mbox{loc}}\, .
	\end{equation}
The interaction   is proposed on the ground of carrying the most important resonances, which define in this case a local polyad $\hat P_{\tiny \mbox{L}}$. As we shall show this is possible because the splitting between the energy levels  is small compared with the energy separation between the fundamentals.
This description is feasible for  water molecule (H$_2$O) for instance, since the masses ratio  between oxygen and hydrogen is 16:1. 

On the other hand, when the molecular masses  are similar  and  a linear geometry is present the normal modes behaviour is preferred with a Hamiltonian of the form
	\begin{equation} 
	\hat{H}_{\tiny \mbox{N}}=\hat{H}_0^{\tiny \mbox{nor}}+\hat{V}_{\tiny \mbox{int}}^{\tiny \mbox{nor}} ,\label{Hnor}
	\end{equation}
where in this case $\hat{H}_0^{\tiny \mbox{nor}}$ represents the contribution of two non-interacting harmonic oscillators associated with the normal modes, while the interaction $\hat{V}_{\tiny \mbox{int}}^{\tiny \mbox{nor}}$ involves diagonal interactions as well as resonant terms that define the  polyad $\hat{P}_{\tiny\mbox{N}}$. This behaviour manifests by a large splitting between the fundamentals. This normal scheme describes with good accuracy the CO$_2$ molecule since the  masses ratio  between carbon and oxygen is 4:3 presenting a linear geometry.

A molecule with a  local mode behaviour may also be described within the normal scheme, with the property  $\hat P_{\tiny\mbox{L}}=\hat P_{\tiny\mbox{N}}$ being satisfied since both descriptions are connected by means of a canonical transformation  \cite{kellman2}. In contrast,  when a  normal mode behaviour is present the polyad $\hat P_{\tiny\mbox{L}}$ stop commuting with the Hamiltonian and $\hat P_{\tiny\mbox{N}} \neq \hat P_{\tiny\mbox{L}}$, as we shall prove in \S2.3.

 In a  vibrational molecular description  a major question is concerned with a criterion to identify  local or normal mode behaviours. To carry out this characterization it has been recently introduced the $\zeta$-parameter as a measure of the locality (normality) degree~\cite{marisol}, in a similar way to the  parameter used in Ref. \cite{child}. This parameter takes into account the splitting of the fundamental modes relative to themselves, i.e.
\begin{equation}
\label{zetap}
\zeta=\frac{2}{\pi}\arctan\left(\frac{\Delta E}{\bar E}\right) \, ,
\end{equation}
 where $\bar E$ denotes the average of fundamental modes and $\Delta E$ the difference between them. We have found the following rough criterion: a value $\zeta < 0.1$ indicates a local character of the molecule while $\zeta>0.1$ a normal one, with a transition region to be identified by means of several concepts \cite{marisol}.


\subsection{Local scheme}

The local description of the stretching degrees of freedom  of a triatomic molecule of type BA$_2$ consists in expanding the Hamiltonian in terms of the local displacement coordinates $q_1=r_1-r_e$ and $q_2=r_2-r_e$, where $r_j$ is  the distance between the central and the $j$-th terminal atom, while $r_e$ is the equilibrium distance of the molecule.  Considering interactions up to quadratic order the Hamiltonian has the form
	\begin{equation}
	\label{Hlocint}
	\hat H={1 \over 2}{g^{{\tiny \mbox{o}}}_{rr}} \sum_{i=1}^2 \hat p_i^2+{1 \over 2}f_{rr} \displaystyle\sum_{i=1}^2 \hat q^2_i	+{g^{{\tiny \mbox{o}}}_{rr'}} \hat p_1 \hat p_2+f_{rr'} \hat q_1 \hat q_2\, ,
	\end{equation}
which should be identified with a Hamiltonian of the form (\ref{Hloc}). In  (\ref{Hlocint}) the parameters  $f_{rr}$ and $f_{rr'}$ correspond to the force constants, while $g_{rr}^{\circ}$ and $g_{rr'}^{\circ}$ are the elements of the Wilson's matrix at equilibrium \cite{gmatrix}, whose explicit expressions in this case are
\begin{align}\label{gs}
	g_{rr}^{\circ}=\frac{1}{m_{{\tiny \mbox{A}}}}+\frac{1}{m_{{\tiny \mbox{B}}}}\,, \qquad 
	g_{rr'}^{\circ}=\frac{\cos{\theta}}{m_{{\tiny \mbox{B}}}}\, , 
	\end{align} \\
where  $\theta$ is the angle between the bonds.  The Hamiltonian \eqref{Hlocint} can be translated into an algebraic representation by introducing the bosonic operators $\hat{a}_i$ and $\hat{a}_i^\dagger$ via the transformation 
	\begin{equation}
	\label{ptoa}
	\hat p_i=i \, \hbar \,\alpha_{{\tiny \mbox{0}}} ~ (  \hat a_i^\dagger-\hat a_i   ), \qquad 
	\hat{q}_i = {1 \over 2 \, \alpha_{{\tiny \mbox{0}}} }  ~( \hat a_i^\dagger+\hat a_i )\, , 
	\end{equation}
with
\begin{equation}
\alpha^2_{{\tiny \mbox{0}}}= {1 \over 2 \hbar} \sqrt{{f_{rr} \over g^o_{rr}}}={  \mu \, \omega_{{\tiny \mbox{0}}}  \over 2 \hbar },
\end{equation}
where the mass-type parameter is $\mu=1/{g^{{\tiny \mbox{o}}}_{rr}}$ and the frequency-type parameter is $\omega_0 =\sqrt{f_{rr}\,{g^{{\tiny \mbox{o}}}_{rr}}}$. Insertion of  the transformation \eqref{ptoa} into the Hamiltonian \eqref{Hlocint}  yields the  algebraic representation
	\begin{equation}\label{hla}
	\hat H = \frac{\hbar \omega_{{\tiny \mbox{o}}}}{2}  \bigg\{ \sum_{i=1}^2 (a_i^\dagger a_i+ a_i  a_i^\dagger)+\lambda  (a_1^\dagger  a_2+ a_1  a_2^\dagger) +\lambda' (a_1^\dagger  a_2^\dagger+ a_1 a_2)  \bigg\}, 
	\end{equation}
with
	\begin{equation}\label{consloca}
	\lambda= \left( 	 x_f + x_g	\right), \qquad
	\lambda'=  \left( x_f - x_g \right) \, 
	\end{equation}
where for convenience we have introduced the definitions
	\begin{equation}\label{xf-xg}
	x_f=\frac{f_{rr'}}{f_{rr}}\, , \qquad
	x_g=\frac{g^{{\tiny \mbox{o}}}_{rr'}}{g^{{\tiny \mbox{o}}}_{rr}}\, .
	\end{equation}
We now define the local polyad as
\begin{equation}
\hat P_L=\hat a_1^\dagger \hat a_1+\hat a_2^\dagger \hat a_2=\hat n_1+\hat n_2.
\end{equation}
\noindent
It is clear that the Hamiltonian (\ref{hla}) does  not commute with $\hat P_L$ unless the last contribution can be neglected. When this is the case the system is identified with a local-mode behaviour with Hamiltonian:
	\begin{equation}\label{ham12poly}
	\hat H_{\tiny \mbox{loc} }
	= \frac{\hbar\,\omega_{{\tiny \mbox{0}}}}{2} \left[ \sum_{i=1}^2 (\hat a_i^\dagger \hat a_i+ \hat a_i \hat a_i^\dagger)
	+ \lambda ~ (\hat a_1^\dagger  \hat a_2+ \hat a_1 \hat a_2^\dagger)\right] .
	\end{equation}
If  anharmonic corrections are added to the diagonal contribution one has a  Morse-like spectrum. Hence, a suitable Hamiltonian to describe the two local oscillators is
	\begin{equation}\label{ham12Morse}
	\hat H_{\tiny \mbox{AOHC} }
	= \hat H^{{\tiny\mbox{M}}}_1+\hat H^{{\tiny\mbox{M}}}_2+
\frac{\hbar\, \omega_0 \,\lambda}{2}(\hat a_1^\dagger  \hat a_2+ \hat a_1 \hat a_2^\dagger), 
	\end{equation}
where $\hat H^{{\tiny\mbox{M}}}_i$ are Morse Hamiltonians. This model is known as the {\it anharmonic oscillators   harmonically coupled} (AOHC)   since the matrix elements of the non diagonal contribution are calculated in the harmonic approximation \cite{lawton,child,halonen,halonen2p,jensen2}. This Hamiltonian is the basis of the local mode theory developed during the 80's.  

\subsection{Normal scheme}

For molecules presenting large splitting between the fundamentals ($\zeta > 0.1$) a normal-mode description   is more convenient. The Hamiltonian (\ref{Hlocint}) can be rewritten in terms of the normal coordinates 
	\begin{equation}
	\hat S_g={1 \over \sqrt{2}}(\hat q_1+\hat q_2)\, , \qquad  
	\hat S_u={1 \over \sqrt{2}}(\hat q_1-\hat q_2)\, ,
	\label{ccn}
	\end{equation}
with the corresponding induced transformation in the momenta
	\begin{equation}
	\hat P_g={1 \over \sqrt{2}}(\hat p_1+\hat p_2), \qquad 
	\hat P_u={1 \over \sqrt{2}}(\hat p_1-\hat p_2)\, .
	\label{pcm}
	\end{equation}
 In these symmetry adapted coordinates the Hamiltonian is diagonal, given as two independent harmonic oscillators (normal modes), i.e.
	\begin{equation}\label{Hnorc}
	\hat H={1 \over 2}( G_{gg} \hat P^2_g+G_{uu} \hat P^2_u)+{1 \over 2}(F_{gg} \hat S_g^2+F_{uu}\hat S_u^2) ,
	\end{equation} 
where 
	\begin{align}
	G_{gg}&=g^{{\tiny \mbox{o}}}_{rr}+g^{{\tiny \mbox{o}}}_{rr'}\, ,\qquad 
	G_{uu}=g^{{\tiny \mbox{o}}}_{rr}-g^{{\tiny \mbox{o}}}_{rr'}\ ,\\
	F_{gg}&=f_{rr}+f_{rr'}\, , \qquad 
	F_{uu}=f_{rr}-f_{rr'}\ . 
	\label{F}
	\end{align}
This is an expected result since the Hamiltonian (\ref{Hlocint}) is integrable. Its corresponding  algebraic representation  is attained by means of the bosonic realization
	\begin{equation}\label{asnor}
	\hat a_{\gamma}^{\dagger }=\frac{1}{\sqrt{2\hbar}}\left(\frac{1}{\alpha_{\gamma_0}}  \hat S_{\gamma}-i\,					\alpha_{\gamma_0}\hat {P}_{\gamma}\right)\, , \qquad
	\hat a_{ \gamma}=\frac{1}{\sqrt{2\hbar}}\left(\frac{1}{\alpha_{\gamma_0}}  \hat S_{\gamma}+i\,							\alpha_{\gamma_0}\hat {P}_{\gamma}\right)\, ,
	\end{equation}
where $\gamma=g,u$, with the definitions 
	\begin{equation}
	\alpha_{g_0}^2=\sqrt{\frac{G_{gg}}{F_{gg}}}={ 1 \over 2 \hbar \alpha^2_{{\tiny \mbox{0}}} r}\,  , \qquad		
	\alpha_{u_0}^2=\sqrt{\frac{G_{uu}}{F_{uu}}}={ 1 \over 2 \hbar \alpha^2_{{\tiny \mbox{0}}} s}\, 
	\end{equation}
where
	\begin{equation}
	r=\sqrt{{1+x_f \over 1+x_g}}, \qquad s=\sqrt{{1-x_f \over 1-x _g}} \, .
	\end{equation}
Hence, in the bosonic space the  Hamiltonian takes the simple form
	\begin{align}\label{hamdnor12}
	\hat H = \frac{\hbar \omega_ {{g}} }{2}   (\hat a_{g}^\dagger \hat a_{g}+ \hat a_{g} \hat a_{g}^\dagger) +\frac{\hbar 			\omega_ {{u}}}{2} (\hat a_{u}^\dagger \hat a_{u}+ \hat a_{u} \hat a_{u}^\dagger)\, ,  
	\end{align}
with frequencies
	\begin{align}\label{wnora}
	\omega_ {{g}}&=\omega_{{\tiny \mbox{0}}} \sqrt{(1+x_f) (1+x_g)}\, , \\
	\label{wnorb}
	\omega_ {{u}}&=\omega_{{\tiny \mbox{0}}} \sqrt{(1-x_f) (1-x_g)}	\, .
	\end{align}
In a normal mode scheme the polyad takes the general form
	\begin{equation}
	\label{polyN}
	\hat P_{\tiny\mbox{N}}= \eta_1 \, \hat n_g+\eta_2 \, \hat n_u,
	\end{equation}
where $\hat{n}_\gamma=\hat a_\gamma^\dagger \, \hat a_\gamma$, while $\{ \eta_1,\eta_2 \}$ are integers fixed  by the resonance. It should be clear that $[ \hat P_{{\tiny \mbox{N}}},\hat H]=0$ and the polyad is preserved. In the next subsection we shall prove that in general $\hat{P}_{{\tiny \mbox{N}}}\neq\hat{P}_{{\tiny \mbox{L}}}$ unless a local mode behaviour is enforced.

\subsection{Local-to-normal connection}

Here we establish the connection between the local scheme \eqref{ham12poly} and the normal  scheme \eqref{hamdnor12}. Taking into account the relation between the local and normal coordinates (\ref{ccn}) as well as the corresponding momenta (\ref{pcm}),  we obtain the connection between the normal and local bosonic operators
\begin{subequations}
\label{exacto}
\begin{eqnarray}
a^\dagger_g&=&{1 \over 2  \sqrt{2r} } \{ (r+1) (a^\dagger_1+ a^\dagger_2)     +   (r-1) (a_1+ a_2)           \}, \\
a^\dagger_u&=&{1 \over 2  \sqrt{2s} } \{ (s+1) (a^\dagger_1- a^\dagger_2)     +   (s-1) (a_1- a_2)           \}, 
\end{eqnarray}
\end{subequations}
Substitution of (\ref{exacto}) into the Hamiltonian (\ref{hamdnor12}) generates the Hamiltonian local representation  (\ref{hla}). We next consider the polyad (\ref{polyN}). The substitution of (\ref{exacto}) into the normal polyad yields
\begin{align}
\label{pnl}
\hat P_{\tiny\mbox{N}}=&\zeta_0+\beta_0  ~ \hat P_{\tiny\mbox{L}}+ \beta_1 ~ (\hat a^\dagger_1 \hat a_2+\hat a_1 \hat a^\dagger_2)+ \nonumber \\
&\beta_2 (\hat a^{\dagger 2}_1 + \hat a^{\dagger 2}_2+ \hat a^{ 2}_1 + \hat a^{ 2}_2)+\beta_3 (\hat a^\dagger_1 \hat a^\dagger_2+\hat a_1 \hat a_2),
\end{align}
where
\begin{subequations}
\label{coef}
\begin{eqnarray}
\zeta_0&=& {2 \over 8 r s}[\eta_1 s(r-1)^2+\eta_2 r (s-1)^2],\\
\beta_0&=& {2 \over 8 r s}[\eta_1 s(r^2+1)+\eta_2 r (s^2+1)],\\
\beta_1&=& {2 \over 8 r s}[\eta_1 s(r^2+1)-\eta_2 r (s^2+1)],\\
\beta_2&=& {1 \over 8 r s}[\eta_1 s(r^2-1)+\eta_2 r (s^2-1)],\\
\beta_3&=& {1 \over 8 r s}[\eta_1 s(r^2-1)-\eta_2 r (s^2-1)].
\end{eqnarray}
\end{subequations}
We have thus proved that in general $\hat P_{\tiny\mbox{N}} \neq \hat P_{\tiny\mbox{L}}$. The question which arises is concerned with the condition under which $\hat P_{\tiny\mbox{L}}=\hat P_{\tiny\mbox{N}}$. This situation is satisfied in the local limit as we next prove. As a first step it is convenient to analyze    the unitary transformation:
	\begin{equation}
	\label{transtoc}
	\hat{a}^\dagger_{g}=\frac{1}{\sqrt{2}} (\hat{c}^\dagger_1+ \hat{c}^\dagger_2)\, , \qquad
	\hat{a}^\dagger_{u}=\frac{1}{\sqrt{2}} (\hat{c}^\dagger_1- \hat{c}^\dagger_2)\, .
	\end{equation}
 When (\ref{transtoc}) is inserted  into the Hamiltonian (\ref{hamdnor12}) one obtains. 
\begin{equation}
\label{hamnorces}
	\hat H =\frac{\hbar \, \omega_{{\tiny \mbox{nor}}}}{2} \sum_{i=1}^2 (c_i^\dagger c_i+ c_i c_i^\dagger)
	+\frac{ \hbar\,\lambda_{{\tiny \mbox{nor}}}}{2} (c_1^\dagger  c_2+ c_1 c_2^\dagger)\,  
\end{equation}
with the definitions
\begin{subequations}
\label{wlnor12}
	\begin{eqnarray}
	\omega_{{\tiny \mbox{nor}}}&=&\frac{\omega}{2} \left( \sqrt{\left(  1+x_f\right)\left(
	1+x_g\right)}+ \sqrt{\left(  1-x_f\right)\left(
	1-x_g\right)}\right),  \\  
	\lambda_{{\tiny \mbox{nor}}}&=&\omega\left( \sqrt{\left(  1+x_f\right)\left(
	1+x_g\right)}- \sqrt{\left(  1-x_f\right)\left(
	1-x_g\right)}\right).
	\end{eqnarray}
\end{subequations}
Notice that the local operators $\hat{c}^\dagger_i$($\hat{c}_i$) do not correspond to the physical local operators $\hat{a}^\dagger_i$($\hat{a}_i$), but their action over an isomorphic local basis  may be chosen to be the same. In fact, we may   establish the isomorphism
	\begin{equation}\label{ctoa}
	\hat{c}^\dagger_i \leftrightarrow \hat{a}^\dagger_i \, , \qquad \hat{c}_i \leftrightarrow \hat{a}_i\, .
	\end{equation}
The difference between the local operators $\hat c^\dagger_i(\hat c_i ) $ and $\hat a^\dagger_i(\hat a_i)$ is manifested by the fact that the coefficients involved in (\ref{ham12poly})  are not equal to the coefficient in (\ref{wlnor12}).  However, a coincidence of the spectroscopic coefficients for molecules with a local mode behaviour is expected.  In order to elucidate the conditions to be satisfied we consider the Taylor series expansion of the parameters (\ref{wlnor12})   as a function of $x_f$ and $x_g$ around zero. The result is
	\begin{align}
	\omega_{{\tiny \mbox{nor}}}&=\frac{ \omega_{{\tiny \mbox{loc}}}}{2} \left( 1+ {1 \over 8} (x_g-x_f)^2+\dots 	{\cal O}(x^3)\right)\, , \\ 
	\lambda_{{\tiny \mbox{nor}}}&= \, \omega_{{\tiny \mbox{loc}}}\left( (x_g+x_f)+{1 \over 8} (x_g^3+x_f^3)-{1 	\over 8} 		(x_g^2 x_f+x_g x_f^2)+ \dots {\cal O}(x^4)\right)\, .
	\label{Taylorx}
	\end{align}
Keeping only  linear terms in $x_f$ and $x_g$ we arrive to the conclusion that with the conditions  
\begin{equation}
\label{lcon}
{1 \over 8}(x_f-x_g)^2\ll 1; \ \ \ \  |x_f| \ll 1; \ \ |x_g| \ll 1,
\end{equation}
we recover the spectroscopic parameters associated with the interacting local oscillators. In other words, in the limit  (\ref{lcon}) the Hamiltonian (\ref{hamdnor12}) reduces to the local representation (\ref{hla})  with the spectroscopic parameters given by $\lambda$ and $\omega_0$,  with $\lambda'=0$. The condition (\ref{lcon}) establishes the local limit, characterized  by the suitability of a local model to estimate the force constants at zeroth order.

In order to see the consequences of the  local condition (\ref{lcon}) in the polyads,  we expand the functions $r$ and $s$ in terms of $x_f$ and $x_g$:
\begin{subequations}
\begin{eqnarray}
r&=&1+{1 \over 2} (x_f-x_g)-{1 \over 8}(x_f+x_g)^2+{x_g^2 \over 2}+\dots \, ,\nonumber \\
s&=&1-{1 \over 2} (x_f-x_g)-{1 \over 8}(x_f+x_g)^2+{x_g^2 \over 2}+\dots \, .
\end{eqnarray}
\end{subequations}
When these expansions are substituted into (\ref{coef}), after applying the local limit we obtain
\begin{equation}
 \beta_0 \approx 1; \qquad   \zeta_0= \beta_i \approx 0 \, ; \qquad i=1,2,3.
\end{equation}
In addition, since in the local limit the interaction between the oscillators produces a small splitting, the equality $\eta_1=\eta_2=1$ is expected and therefore one gets
\begin{equation}
 \hat P_{\tiny\mbox{N}} \approx \hat P_{\tiny\mbox{L}}.
\end{equation}
Simultaneously   the Bogoliubov-type transformation (\ref{exacto}) reduces to the canonical transformation (\ref{transtoc}).

This analysis  shows that starting with a local mode behaviour there is a transition region where the equality $\hat P_{\tiny\mbox{L}}=\hat P_{\tiny\mbox{N}}$ stop being valid as well as the   estimation of the correct force constants at zeroth order in a local model. This transition may be simulated by changing the strength of the interaction in the kinetic energy, which may be interpreted as modification in the mass ratio of the atoms \cite{terasaka}, but also through the whole transformation from a molecular system to another one \cite{marisol}. In this contribution we shall consider  this local-to-normal transition via a time dependent excitations in the sudden, linear and adiabatic form by modifying the inter-bond angle as a simulation of an electronic state transition. In particular we   are interested in studying  the system as a  transient LNT involving  the polyad breaking point of view. 


\section{Time-dependent description of vibrational states}

In this contribution we study the  energy level pattern of the mean value of the Hamiltonian provided by the stretching degrees of freedom of triatomic molecules
 when a geometrical change  takes place via a time-dependent excitation through the angle $\theta(t)$ between the bonds. Hence, with this  geometric dependence a molecule with local character may move to a normal mode  behaviour and vice versa. From the physical point of view this situation may be considered as a simulation of an electronic state change.
To achieve this goal we start by  describing the evolution of the states by considering the Hamiltonian  \eqref{Hnorc} in terms of symmetry adapted coordinates, where now the parameters $G_{gg}$ and $G_{uu}$ become  time-dependent functions given by 
	\begin{equation}\label{Gs}
	G_{gg}(t) = \frac{1}{\mu} + \frac{ \cos{\theta(t)} }{ m_B }\, ,  
	\qquad G_{uu}(t) = \frac{1}{\mu} + \frac{ \cos{\theta(t)} }{ m_B }\ . 
	\end{equation} 
Concerning this geometrical time dependence, it is important to mention that from a physical point of view this change implies a modification of the potential energy surface and consequently a time dependence in the parameters $F_{gg}(t)$  and $F_{uu}(t)$  should in principle be considered.

\subsection{Normal scheme}

The time-evolution of the vibrational  Hamiltonian \eqref{Hnorc} consists in the  two independent parametric oscillators, i.e.,  
	\begin{equation}\label{POH}
    	\hat{H}=\frac{1}{2}\left\{ G_{gg}(t)\hat{P}_g^2+F_{gg}(t)\hat{S}_g^2+ G_{uu}(t)\hat{P}_u^2+F_{uu}(t)		\hat{S}_u^2\right\} \, ,
    	\end{equation}
where the functions $G_{gg}(t)$ and $G_{uu}(t)$ are given by \eqref{Gs}, while the potential parameters $F_{gg}(t)$ and $F_{uu}(t)$  depend on the electronic dynamics. The solutions of the time-dependent Schr\"odinger equation corresponding to the above Hamiltonian may be obtained by means of the invariant operators method~\cite{lewis, malkin1970coherent, malkin1973linear, Hans}, as we next show.

 In the Heisenberg picture an operator $\hat{I}(t)$ is said to be invariant when its total time derivative vanishes, i.e.
	\begin{equation} 
	\frac{d \hat I}{dt}= \frac{i}{\hbar}\left[\hat H, \hat I \right] + \frac{\partial \hat I}{\partial t} = 0\ .
	\end{equation}
In  equivalent form, in the Schr\"odinger picture an operator $\hat{I}(t)$ is  invariant when the  mean value with respect to any state is time-independent. When this is the case  it is said that $\langle \hat I \rangle$ is a constant of motion of the system.
 In particular, a parametric oscillator system possesses an invariant known as the Ermakov operator~\cite{schuch2007connection, schuch2008riccati,Hans}, and  due to the fact that  the Hamiltonian \eqref{POH} consists of the sum of two uncoupled parametric oscillators, its invariant takes the form: $\hat I (t) = \hat{I}_g(t) + \hat{I}_u(t)$, where $ \hat{I}_g(t)$ and $ \hat{I}_u(t)$ are Ermakov operators given by
    	\begin{equation}\label{Ermakov-invariant}
    	\hat{I}_{\gamma}(t) 
	=\frac{1}{2\hbar}\left[ 
    	\left( \alpha_{\gamma}(t)\hat{P}_{\gamma}-\frac{\dot{\alpha}_{\gamma}(t)}{G_{\gamma\gamma}(t)}\hat{S}_{\gamma}
    	\right)^2+\left(\frac{\hat{S}_{\gamma}}{\alpha_{\gamma}(t)} \right)^2
    	\right]\, ; \qquad \gamma=g,\,u.
    	\end{equation}
Here each operator by itself is also  an invariant of \eqref{POH} and  the time-dependent functions $\alpha_\gamma(t)$ are solutions to the non-linear Ermakov-type equation
    \begin{equation}\label{Ermakov-equation}
    \ddot{\alpha}_\gamma-\frac{\dot{G}_{\gamma\gamma}(t)}{G_{\gamma\gamma}(t)}\dot{\alpha}_\gamma+G_{\gamma\gamma}(t)F_{\gamma\gamma}(t)\alpha_\gamma=\frac{G_{\gamma\gamma}^2(t)}{\alpha_\gamma^3(t)}\,.
    \end{equation}

The solutions of the Schr\"odinger equation associated with the Hamiltonian \eqref{POH} are determined by the eigenvectors of the invariant $\hat I_\gamma(t)$~\cite{lewis, malkin1973linear, Hans}. In order  to find such eigenvectors we should notice that each invariant $I_\gamma(t)$ can be expressed as  
\begin{equation}
\label{Ig}
\hat{I}_\gamma(t)=\hat{A}_\gamma^\dagger(t)\hat{A}_\gamma(t)+1/2,
\end{equation}
where the operators $\hat{A}_\gamma^\dagger(t)$ and $\hat{A}_\gamma(t)$ are also invariant operators given by
	\begin{subequations}
	\label{ast}
	\begin{eqnarray}
	\hat{A}_\gamma(t)&=\frac{e^{i\phi_\gamma(t)}}{\sqrt{2\,\hbar}}\left[ \left(\frac{1}{\alpha_\gamma(t)} -i\frac{\dot{\alpha_		\gamma}(t)}{G_{\gamma\gamma}(t)} \right)\hat{S}_\gamma+i\alpha_\gamma(t)\hat{P}_\gamma \right],\label{asc-op}\\
	\hat{A}_\gamma^\dagger(t)&=\frac{e^{-i\phi_\gamma(t)}}{\sqrt{2\,\hbar}}\left[ \left(\frac{1}{\alpha_\gamma(t)} +i				\frac{\dot{\alpha}_\gamma(t)}{G_{\gamma\gamma}(t)} \right)\hat{S}_\gamma-i\alpha_\gamma(t)\hat{P}_\gamma \right]\, .		\label{desc-op}	
	\end{eqnarray}
\end{subequations}
with  phase factor $\phi_\gamma(t)=\int^t\frac{G_{\gamma\gamma}(\tau)}{\alpha_\gamma^2(t)}d\tau$~\cite{SOO}.
The  linear operators (\ref{ast}) obey the commutation relations:
\begin{equation}
[ \hat{A}_{\gamma'}(t), \hat{A}_\gamma^\dagger(t)]=\delta_{\gamma \gamma'}; \qquad [ \hat{A}_{\gamma'}(t), \hat{A}_\gamma(t)]=[ \hat{A}^\dagger_{\gamma'}(t), \hat{A}_\gamma^\dagger(t)]=0,
\end{equation}
and consequently  they possess the same algebraic properties as the bosonic operators associated with the harmonic oscillator  corresponding to the well-known Heisenberg--Weyl group $H_4$~\cite{zhang1990coherent}. Hence  the Hilbert space can be spanned by the time-dependent Fock states $|n_\gamma,t\rangle$
	\begin{equation}\label{Fock-state}
	|n_\gamma,t\rangle=\frac{1}{\sqrt{n_\gamma\,!}}\left[\hat{A}_\gamma^\dagger(t)\right]^{n_\gamma}| 0_\gamma , t 		\rangle\, .
	\end{equation}
The eigenvectors of the Ermakov operator $\hat I_\gamma(t)$ are then the  states $|n_\gamma,t\rangle$ with eigenvalues $n_\gamma+1/2$. Taking the direct product $| n_g,n_u,t\rangle = | n_g , t \rangle \otimes |n_u,t\rangle$, we have for the eigensystem associated with $\hat{I}(t)$:  
	\begin{equation}
	\hat{I}(t)| n_g,n_u,t\rangle =\left( n_g + n_u + 1 \right)| n_g,n_u,t\rangle.
	\end{equation}
The expression for the ground states $| 0_\gamma, t \rangle$ in the position representation is attained by solving the partial differential equation defined by the equation $\langle S_\gamma | \hat {A}_\gamma | 0_\gamma , t \rangle~=~0$, whose normalized solution is given by
{\small{
	\begin{equation}
	\psi_{0_\gamma}(S_\gamma,t)= \frac{1}{(\pi\hbar)^{1/4}}\frac{1}{\sqrt{\alpha_\gamma(t)}}\exp\Bigg\{ \frac{i}{2\,\hbar\, 		G_{\gamma\gamma}(t)}\left(\frac{\dot{\alpha}_\gamma(t)}	{\alpha_\gamma(t)}+i\frac{G_{\gamma\gamma}(t)}				{\alpha_\gamma^2(t)}\right)S_\gamma^2-\frac{i}{2}\int^t\frac{G_{\gamma\gamma}(\tau)}{\alpha_\gamma^2(\tau)}d\tau		\Bigg\}\, .
	\end{equation} 
}}
On the other hand, from the ground state $\psi_{0_\gamma}(S_\gamma,t)$  and the definition \eqref{Fock-state} it is possible to find the wave function for an arbitrary Fock state through successive applications of the creation operator $\hat{A}_\gamma^\dagger(t)$,  wherefore one obtains that
	\begin{equation}
\label{WFG}
	\psi_{n_\gamma}(S_\gamma,t) = \frac{\psi_{0_\gamma}(S_\gamma, t)}{\sqrt{2^{n_\gamma}{n_\gamma}!}}H_{n_\gamma}\left( \frac{1}		{\sqrt{\hbar}}\frac{S_\gamma}{\alpha_\gamma(t)}\right)e^{i \, n_\gamma \, \phi_\gamma(t)}\, .
	\end{equation}
Finally, by direct substitution it is not difficult to  prove that the wave functions 
	\begin{equation}
	\label{WF}
	\Psi_{n_g n_u}(S_g, S_u, t)=\psi_{n_g}(S_g,t)\,\psi_{n_u}(S_u,t)
	\end{equation}
are solutions of the time-dependent  Schr\"odinger equation associated with the Hamiltonian \eqref{POH}.  

To complete the study of the system dynamics,  we include the evolution of the quantum uncertainties for each parametric oscillator, defined as 
	\begin{equation}
	\sigma_{S_\gamma}^2(t)=\langle \hat{S}_\gamma^2 \rangle - \langle \hat{S}_\gamma \rangle^2\, , \qquad
	\sigma_{P_\gamma}^2(t)=\langle \hat{P}_\gamma^2 \rangle - \langle \hat{P}_\gamma \rangle^2\, ,
	\end{equation} 
as well as the evolution of their correlation, 
	\begin{equation}
	\sigma_{S_\gamma P_\gamma}(t)=\frac{1}{2}\langle \hat{S}_\gamma \hat{P}_\gamma+\hat{P}_\gamma\hat{S}_			\gamma \rangle - \langle \hat{S}_\gamma \rangle \langle \hat{P}_\gamma \rangle\, ,
	\end{equation}
which can  be determined by direct computation of the mean values with respect to the wave functions (\ref{WFG}), a task that yields
	\begin{eqnarray}
	\sigma^2_{S_\gamma}(t)&=&\frac{h}{2}\,\alpha_\gamma^2(t)(2\,n_\gamma+1)\, ,
	\label{sigma-x}\\
	\sigma^2_{P_\gamma}(t)&=&\frac{h}{2}\,\left[ \left( \frac{\dot{\alpha}_\gamma(t)}{G_{\gamma\gamma}(t)} \right)^2+			\frac{1}{\alpha_\gamma^2(t)}\right](2\,n_\gamma+1)\, ,
	\label{sigma-p}\\
	\sigma_{S_\gamma P_\gamma}(t)&=&\frac{h}{2}\,\frac{1}{G_{\gamma\gamma}(t)}\,\dot{\alpha}_\gamma(t)\,\alpha_			\gamma(t)(2\,n_\gamma+1)\, .
	\label{sigma-xp}
	\end{eqnarray}
 Moreover, since the mean value of the Hamiltonian \eqref{POH} has the form
	\begin{equation}\label{spectrum}
	\langle \, \hat{ H} \, \rangle(t)= \sum_{\gamma = g,\, u} \frac{G_{\gamma\gamma}(t)}{2}\sigma^2_{P_\gamma}(t)+			\frac{F_{\gamma\gamma}(t)}{2}\sigma^2_{S_\gamma}(t)\, ,
	\end{equation}
and consequently it can be computed with the help of expressions \eqref{sigma-x} and \eqref{sigma-p}. 
Therefore, from this analysis we see that the dynamics of the quantum properties is completely determined by the solutions of the nonlinear Ermakov equation \eqref{Ermakov-equation} with the initial conditions 
	\begin{equation}
	\alpha_\gamma(t_0) \equiv \alpha_{0_\gamma}=\left(\frac{G_{\gamma\gamma}(t_0)}{F_{\gamma\gamma}		(t_0)}\right)^{\frac{1}	{4}}\, \qquad 
	\dot{\alpha}_\gamma(t_0) \equiv \dot{\alpha}_{0_\gamma}=0\, .
	\end{equation}
 From this initial condition, we notice that the solution $\alpha_\gamma (t)$  has dimensions of $[time/mass]^{-1/2}$, which in practice is taken to be $[fs/amu]^{-1/2}$.	
	
In this time-dependent description  the normal polyad defined by $\hat{P}_{{\tiny \mbox{N}}} =\eta_1\, \hat{n}_g + \eta_2\,\hat{n}_u$ is not  a good quantum number anymore since  it is not an invariant operator for the system. Indeed, the time evolution for its mean value takes the form
\begin{equation}
	\langle \hat P_{{\tiny \mbox{N}}} \rangle(t)=\frac{1}{2\hbar}\sum_{\gamma = g,\, u}  \eta_i \left[ \alpha_{\gamma_0}^2 			\sigma^2_{P_\gamma}(t)	+\frac{1}{\alpha_{\gamma_0}^2}\sigma^2_{S_\gamma}(t)\right] - 1\, ,
	\end{equation}
which is a time-dependent quantity.  However, if  we now define the time-dependent normal polyad, $\hat{\mathscr{P}}_{{\tiny \mbox{N}}}(t)$, in the following form	
	\begin{equation}
\label{Pnor}
	\hat{\mathscr{P}}_{{\tiny \mbox{N}}}(t) =\eta_1 \,  \hat{A}_g^\dagger(t) \hat{A}_g (t) + \eta_2 \, \hat{A}_u^\dagger(t) 			\hat{A}_u(t) 
	= \eta_1\, \hat{N}_g(t) + \eta_2 \,  \hat{N}_u(t) 	 
	\end{equation}
where as in the stationary case the integers  $\{ \eta_1,\eta_2 \}$ are fixed by the resonance at initial time, we have by construction  an invariant operator with mean value $\langle \hat{\mathscr{P}}_{{\tiny \mbox{N}}} \rangle = \eta_1\,n_g + \eta_2\,n_u$. At the  initial time $t=t_0$ the polyad (\ref{Pnor}) coincides to the stationary normal polyad, i.e. $\hat{\mathscr{P}}_{{\tiny \mbox{N}}}(t_0)=\hat{P}_{{\tiny \mbox{N}}}$, since $\hat{A}^\dagger_\gamma(t_0)=\hat{a}^\dagger_\gamma$ and $\hat{A}_\gamma(t_0)=\hat{a}_\gamma$.


\subsection{Local scheme} 

Let us now turn our attention to the molecular description in terms of local coordinates. The time-dependent Hamiltonians \eqref{POH} reads 
	\begin{equation}
	\label{TD-local}
	\hat H={1 \over 2}{g^{{\tiny \mbox{o}}}_{rr}} \sum_{i=1}^2 \hat p_i^2+{1 \over 2}f_{rr}(t) \displaystyle\sum_{i=1}^2 \hat 	q^2_i+{g^{{\tiny \mbox{o}}}_{rr'}} (t)\hat p_1 \hat p_2+f_{rr'}(t) \hat q_1 \hat q_2\, .
	\end{equation}
Since the connection between the local and normal coordinates is given by the canonical transformation (\ref{ccn}), the operators $\hat{A}_\gamma(t)$ and $\hat{A}^\dagger_\gamma(t)$, -- as well as the Ermakov invariant $\hat I(t)$ and the polyad $\hat{\mathscr{P}}_{{\tiny \mbox{N}}}(t)$ -- are invariant operators of the Hamiltonian \eqref{TD-local}. These operators when  written in terms  of the local coordinates $(\hat{q}_j, \hat{p}_j)$ take the form
	\begin{equation}
\label{ag12}
	\hat{A}_g (t)=\frac{1}{\sqrt{2}}\left[ \hat{A}_{1 g}(t) + \hat{A}_{2 g}(t) \right]\, , \quad 
	\hat{A}_u (t)=\frac{1}{\sqrt{2}}\left[ \hat{A}_{1 u}(t) - \hat{A}_{2 u}(t) \right]\, , 
	\end{equation}
with the local bosonic operators $\hat{A}^\dagger_{j \gamma}(t)$ and $\hat{A}_{j \gamma}(t)$ given by 
	\begin{eqnarray}
\label{ajg}
	\hat{A}_{j \gamma}(t)&=&\frac{e^{i\phi_\gamma(t)}}{\sqrt{2\,\hbar}}\left[ \left(\frac{1}{\alpha_\gamma(t)} -i				\frac{\dot{\alpha_\gamma}(t)}{G_{\gamma\gamma}(t)} \right)\, \hat{q}_j + i\alpha_\gamma(t)\, \hat{p}_j \right]\, , \\
\label{aju}
	\hat{A}^\dagger_{j \gamma}(t)&=&\frac{e^{- i\phi_\gamma(t)}}{\sqrt{2\,\hbar}}\left[ \left(\frac{1}{\alpha_\gamma(t)} + i		\frac{\dot{\alpha_\gamma}(t)}{G_{\gamma\gamma}(t)} \right)\, \hat{q}_j - i\alpha_\gamma(t)\, \hat{p}_j \right]\, ,
	\end{eqnarray}
where $j=1,\, 2$. It is worth noticing  that the operators $\hat{A}_{j \gamma}(t)$ and $\hat{A}^\dagger_{j \gamma}(t)$ by themselves are not invariant operators of \eqref{TD-local}.

To analyze the local limit (\ref{lcon}) for the time-dependent case, we shall consider the conditions (\ref{lcon}). These conditions will help us to elucidate how the local-to-normal and the normal-to-local transitions take place, as we shall see below.  

From Ermakov-type equation \eqref{Ermakov-equation} it is straightforward to show that in the local limit \eqref{lcon} the solutions $\alpha_g(t)$ and $\alpha_u(t)$ coincide, i.e. $\alpha_g(t)=\alpha_u(t)=\alpha(t)$, and both become  solution to the Ermakov equation
	\begin{equation}
\label{alfal}
	\ddot{\alpha}+\omega^2(t)\alpha = \frac{1}{\mu^2 \alpha^3}\, ,
	\end{equation} 
with initial conditions
	\begin{equation}
	\alpha(t_0)\equiv\alpha_0=\frac{1}{\sqrt{\mu \, \omega(t_0)}}\, , \qquad \dot{\alpha}(t_0)=\dot{\alpha}_0 = 0\, ,
	\end{equation}
considering the time-dependent frequency $\omega(t)=\sqrt{ g^{{\tiny \mbox{o}}}_{rr} f_{rr}(t)}$ and the reduced mass $\mu~=~1/g^{{\tiny \mbox{o}}}_{rr}$. Moreover, it is not difficult to see that $\hat{A}_{1 g}(t)=\hat{A}_{1 u}(t)=\hat{A}_{1}(t)$ and $\hat{A}_{2 g}(t)=\hat{A}_{2 u}(t)=\hat{A}_{2}(t)$. Therefore the connection between normal and local bosonic operators become 
	\begin{equation}
\label{a12gu}
	\hat{A}_g (t)=\frac{1}{\sqrt{2}}\left[ \hat{A}_{1}(t) + \hat{A}_{2}(t) \right]\, , \quad 
	\hat{A}_u (t)=\frac{1}{\sqrt{2}}\left[ \hat{A}_{1}(t) - \hat{A}_{2}(t) \right]\ ,
	\end{equation}
where
\begin{eqnarray}
\label{qjpj}
	\hat{A}_{j}(t)&=&\frac{e^{i\phi(t)}}{\sqrt{2\,\hbar}}\left[ \left(\frac{1}{\alpha(t)} -i				\frac{\dot{\alpha}(t)}{\mu} \right)\, \hat{q}_j + i\alpha(t)\, \hat{p}_j \right]\, , 
	\end{eqnarray}
with $j=1,\, 2$, and $\alpha(t) $ is solution of (\ref{alfal}). From (\ref{ag12}-\ref{aju}) and (\ref{a12gu}-\ref{qjpj}) the connection between the local operators $\hat a^\dagger_i (\hat a_i)$ and the time dependent operators $\hat A^\dagger_i(t) (\hat A_i(t))$  can be established. This correspondence is shown in the Appendix~\ref{Appendix}.

The Ermakov operator in local coordinates takes the form
	\begin{equation}
	\hat{I}_{{\tiny \mbox{L}}}(t) = \hat{A}^\dagger_1(t)\hat{A}_1(t) + \hat{A}^\dagger_2(t)\hat{A}_2(t) + 1,
	\end{equation} 
with eigenstates $|n_1, n_2, t \rangle = |n_1, t\rangle \otimes |n_2, t \rangle$ and corresponding eigenvalues: $n_1 + n_2 +1$. 
 The 
wave functions of the Schr\"odinger equation in the position representation 
are given by 
	\begin{equation}
	\Psi_{n_1 n_2}(q_1, q_2, t)=\psi_{n_1}(q_1,t)\,\psi_{n_2}(q_2,t).
	\end{equation}
The ground state of these functions are determined by
\begin{equation}
	\psi_{0_j}(q_j,t)= \frac{1}{(\pi\hbar)^{1/4}}\frac{1}{\sqrt{\alpha(t)}}\exp\Bigg\{ \frac{i \mu}{2\,\hbar}\left(\frac{\dot{\alpha}		_(t)}{\alpha(t)}+i\frac{1}{\mu\,\alpha^2(t)}\right)q_j^2-\frac{i}{2}\int^t\frac{d\tau}{\mu\,\alpha^2(\tau)}\Bigg\}\, ,
	\end{equation}
while for the wave functions $\psi_{n_j}(q_j,t)$:
	\begin{equation}
	\psi_{n_j}(q_j,t) = \frac{\psi_{0_j}(q_j, t)}{\sqrt{2^{n_j}{n_j}!}}H_{n_j}\left( \frac{1}{\sqrt{\hbar}}\frac{q_j}{\alpha(t)}\right)e^{i \, n_j \, \phi(t)}\, .
	\end{equation}

The time evolution of the mean values of the local polyad  $\hat{P}_{\tiny \mbox{L}}=\hat{n}_1 + \hat{n}_2$,  has in general the time-dependent form
	\begin{equation}\label{Local-Polyad}
	\langle \hat{P}_{\tiny \mbox{L}} \rangle(t) = \frac{1}{2\hbar}\sum_{\gamma=g,u}\left[ \mu \omega \sigma^2_{S_\gamma}(t) 
	+\frac{1}{m \omega} \sigma^2_{P_\gamma}(t) \right]	- 1
	\end{equation}
whereby it is straightforward to see that it  is not  a preserved quantity and consequently it does not provide  a good quantum number. However, the appearance of the local operators $\hat{A}_1(t)$ and $\hat{A}_2(t)$ suggests the  introduction of the time-dependent local polyad operator as 
	\begin{equation}
	\label{Ploc}
	\hat{\mathscr{P}}_{{\tiny \mbox{L}}}(t) = \hat{A}_1^\dagger(t) \hat{A}_1 (t) + \hat{A}_2^\dagger(t) \hat{A}_2(t) 
				          = \hat{N}_1(t) + \hat{N}_2(t) 
	\end{equation}
which is preserved in time in the local limit and whose mean value corresponds to $\langle \hat{\mathscr{P}}_{{\tiny \mbox{L}}} \rangle = n_1 + n_2$. At initial time $t_0$ the local polyad (\ref{Ploc}) coincides with the stationary local polyad $\hat{P}_{\tiny \mbox{L}}$, i.e. $\hat{\mathscr{P}}_{{\tiny \mbox{L}}}(t_0)=\hat{P}_{{\tiny \mbox{L}}}$. The connection between normal and the local time-dependent polyads is via the local limit (\ref{lcon})
\begin{equation}
 	\lim_{|x(t)|<<1}\hat{\mathscr{P}}_{{\tiny \mbox{N}}}(t) = \hat{\mathscr{P}}_{{\tiny \mbox{L}}}(t)\, .
	\end{equation}

We may also consider  the inverse process, i.e. when the evolution of a local molecule evolves with an  increasing functions of time for the parameters 
 $x_f(t)$ and $x_g(t)$. At the beginning of the evolution the states can be labeled by the $\hat{\mathscr{P}}_{{\tiny \mbox{L}}}(t)$. However, when the conditions  \eqref{lcon} are not  satisfied  the time-dependent local polyad is not  a good label anymore. This is the time-dependent version of the polyad breaking phenomenon discussed in \S2.3.



\subsection{Time-dependent molecular behaviour}

It has been shown that the time-dependent mode behaviour of the molecule  is strongly related to the evolution to the functions $x_f(t)$ and $x_g(t)$. It is important however to quantitatively characterize the behaviour of the molecule. To achieve this goal we consider the time evolution of the $\zeta$-parameter defined in \eqref{zetap}, i.e 
	\begin{equation}
	\label{zetat}
	\zeta(t)=\frac{2}{\pi}\arctan\left(\frac{\Delta E(t)}{\bar{E}(t)}\right),
	\end{equation}
where 
	\begin{align}
	\Delta E (t)=|\langle H \rangle _{\nu_g}(t)-\langle H \rangle _{\nu_u}(t) |\, , \qquad
	\bar{ E}(t)=\frac{\langle H \rangle _{\nu_g}(t) + \langle H \rangle _{\nu_u}(t) }{2}\, ,
	\end{align}
 being $\langle H \rangle _{\nu_\gamma}(t)$ the energy of the fundamental modes. Therefore, the time-dependent parameter $\zeta(t)$ characterizes the time-dependent evolution of the molecular behaviour. When  this  parameter decreases  the splitting between states of the same polyad  decreases  going to degeneracy, which implies that  both $\alpha_g (t)$ and $\alpha_u (t)$ simultaneously approach to $\alpha(t)$:   $\hat{\mathscr{P}}_{{\tiny \mbox{N}}} \to $  $\hat{\mathscr{P}}_{{\tiny \mbox{L}}}$. When the parameter $\zeta$  increases the splitting   between states increases too. 

The time dependent change in the angle between bonds, $\theta(t)$, can produce either local-to-normal or normal-to-local mode transition depending on the geometrical and/or mass change, since $\theta(t)$ can modify the functions $x_f(t)$ and $x_g(t)$. Some examples are given in the next section.

\section{Local-to-normal and normal-to-local  mode transitions}

Along the previous discussion  it has been established the conditions under which  either a local-to-normal  or a normal-to-local mode transition takes place for triatomic molecules via a time-dependent excitation through a geometrical change $\theta(t)$. In this section we study these kind of transitions for two local molecules, H$_2$O and O$_3$, going to a normal mode behaviour, and for two normal molecules, CO$_2$ and NO$_2$, going to a local mode behaviour.  These molecules have been classified by means of the $\zeta$-parameter shown in table \ref{t1}.The spectroscopic parameters of these molecules are also displayed in  Table~\ref{t1}. 

The transitions are performed in  sudden, linear and adiabatic form, established by


 \begin{table}
	\tbl{Spectroscopic parameters of molecules A$_2$B type}
  {  \begin{tabular}{|c|c|c|c|c|c|c|c|c|c|c|c|}
	\hline 
    	 	& $f_{rr}$$^a$ & $f_{rr'}$$^a$ & $x_f$ &$g_{rr}^{\circ}$$^b$ &$g_{rr'}^{\circ} $$^b$ & $x_g$ & $\theta_0$ & $\theta_f$ &  $E_{\nu_1}$$^c$ &  $E_{\nu_3}$$^c$ &  $\zeta$ \\
    	 		\hline 
 	CO$_2$~\cite{LemusCO2}  & 15.97 & 1.232 & 0.077 & 0.1458 & -0.083 & -0.571 & 180   & 104.5 & 1285.4 & 2349.1 & 0.337 \\
 		\hline 
	NO$_2$~\cite{frrNO2}& 10.91 & 1.935 & 0.177 & 0.1339 & -0.050 & -0.373 & 134.3 & 104.5 & 1319.8 & 1619  & 0.128 \\
		\hline 
	O$_3$~\cite{frrO3}& 6.164 & 1.603 & 0.260 & 0.125 & -0.028 & -0.225 & 116.8 & 180    & 1104.3 & 1038.7 & 0.039 \\
		\hline 
	H$_2$O~\cite{Lemusvib}& 8.093 & -0.157 & -0.019 & 1.063 & -0.016 & -0.015 & 104.5 & 180  & 3657.1 & 3755.9 & 0.017 \\
		\hline 
    \end{tabular}}%
	\tabnote{$^{\rm a}[\mbox{aJ}\AA^{-2}$}] \tabnote{$^{\rm b}[\mbox{amu}^{-1}$]}\tabnote{$^{\rm c}[\mbox{cm}^{-1}$]}
	\label{t1}
\end{table}%
	
	\begin{equation}\label{transf-1}
	\theta_{{\tiny \mbox{sud}}}(t)=
		\begin{cases}
      			\theta_0 &   t < t_0\\
      			\theta_f  &    t_0 \leq t
    		\end{cases}
		\, ,
	\end{equation}
	\begin{equation}\label{transf-2}
	\theta_{{\tiny \mbox{lin}}}(t)=
		\begin{cases}
			\theta_0 & t < t_0\\
			\frac{\theta_f \, - \, \theta_0}{t_f - t_0} \, (t - t_0)+\theta_0 & t_0\leq t < t_f \\
      	 		\theta_f &  t_f \leq t
    		\end{cases}
		\, ,		 
	\end{equation}

	\begin{equation}\label{transf-3}
	\theta_{{\tiny \mbox{adi}}}(t)= \theta_0+\frac{\theta_f-\theta_0}{1+2\,e^{ - 2\,k\,t}}\, ,
	\end{equation}	
respectively. In all the cases the angle goes from the initial value $\theta_0$ at $t_0$ time, to the final value $\theta_f$ at $t_f$ time. In particular, for the adiabatic case the initial angle $\theta_0$ takes place at $t_0 = -\infty$ while the final angle $\theta_f$  at $t_f = \infty$. Besides in transformation (\ref{transf-3}) the parameter $k$ modulates the angle change speed.
For the linear and the adiabatic cases all the quantum properties are obtained using the numerical solution of the Ermakov equation \eqref{Ermakov-equation}, with the spectroscopic  parameters given in  Table~\ref{t1}.  In the sudden case as one goes from an initial Hamiltonian $\hat{H}_0$ to a final Hamiltonian $\hat{H}_f$, and the evolution of the quantum observables are obtained taking into account that before the change,  the states are  eigenstates of the initial Hamiltonian $\hat{H}_0$ and immediately after the  change the same  states evolved in the final Hamiltonian $\hat{H}_f$, as discussed in  Ref.~\cite{messiah}.  For the examples developed here we consider for the sake of simplification  that the parameters $F_{\gamma \gamma}(t)$ with  $\gamma=\{u,g\}$,  are constants taking their initial value at $F_{\gamma \gamma}(t_0)$.

\subsection{Local-to-Normal mode transition}

Water and ozone molecules present a strong local-mode behaviour. For these molecules we have performed a time-dependent change in their geometry, going from a bending geometry with $\theta_0$ to a linear geometry with $\theta_f = 180^{\circ}$, as displayed in Table~\ref{t1}. So, by means of the solution of the Ermakov equation $\alpha_g(t)$ and $\alpha_u(t)$ the position and momentum uncertainties   were constructed from expressions \eqref{sigma-x} and \eqref{sigma-p} and  via the result \eqref{spectrum} the evolution of the mean value of the Hamiltonian is obtained.

	\begin{figure}[h!]
		\begin{center}
		\setlength{\unitlength}{1pt}
			\includegraphics[width = 13.5 cm]{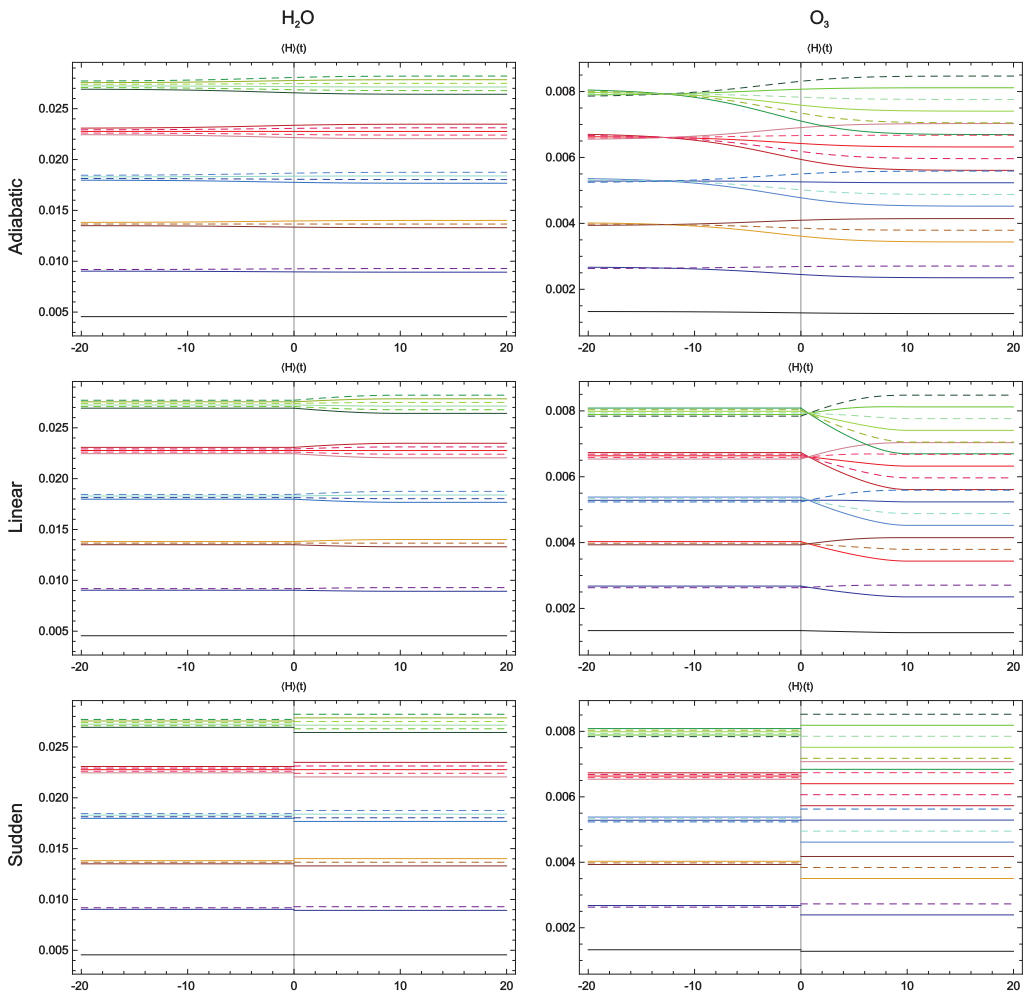}
			\caption{Evolution of the mean value of the Hamiltonian in  units of $\mbox{amu}\AA^2 /fs$, while the time is measure in $fs$ (femtoseconds) and the mass in $amu$(atomic mass units). The continuous lines stand for the gerade states and dashed lines for the ungerade.} 
			\label{Fig-1}
		\end{center}
	\end{figure}

Figure~\ref{Fig-1} displays the time evolution of the mean values of the Hamiltonian \eqref{Hnorc}, $\langle \hat H \rangle(t)$, for the molecules of interest. Sudden, linear and adiabatic changes are presented. As  expected, for times before  $t_0$ the molecular states   are grouped by polyads, i.e. states of the same polyad have small splitting between them. 

For the  water molecule, left panels of Figure~\ref{Fig-1}, as soon as time evolution  starts the splitting between the states belonging to a given  polyad increases. When the molecule reaches the linear geometry at $t_f$ the splitting remains constant. In this system the angle change does not  induce a local-to-normal mode behavior  as manifested by the local polyad preserving mean values of the Hamiltonian.

For ozone molecule  O$_3$, right panels in Figure~\ref{Fig-1},  the splitting of the states grouped by polyad decrease rapidly and collapse into a single state, but  later on  they are reorganized and the splitting increases reaching  constant values. The collapse to a harmonic oscillator degenerate levels is due to the cancellation of the $(x_g+x_f)$ term. It becomes clear that in this case a geometry change  induces a normal mode behaviour, which is explained by the unit mass ratio. Here the strong interaction between states of different polyad become apparent. A polyad breaking is present manifested by the overlapping of the polyads. 

	\begin{figure}[ h ! ]
		\begin{center}
		\setlength{\unitlength}{1pt}
			\includegraphics[width = 8 cm]{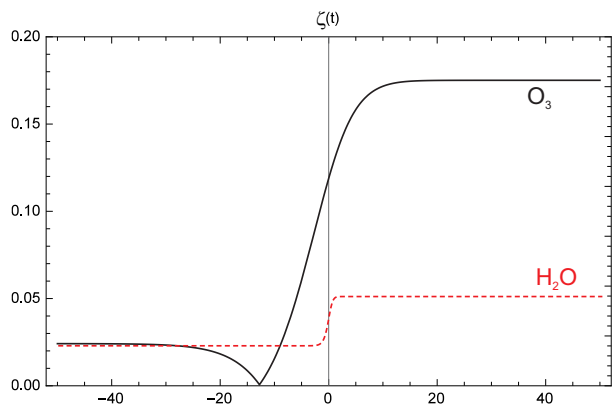}
			\caption{$\zeta(t)$ function for local molecules. Only the  adiabatic change is shown.} 
			\label{Fig-2}
		\end{center}
	\end{figure}

In Figure~\ref{Fig-2} the $\zeta(t)$ parameter  for the two molecules  H$_2$O and O$_3$ in the adiabatic case is displayed.  In both cases the function $\zeta(t)$  increases  as  time evolves. For water molecule the final value $\zeta(t_f)=0.050$ indicates that this molecule does not change its local behavior. Conversely, the ozone final value $\zeta(t_f)=0.174$ reflects the normal behaviour of this molecule after the excitation. This final values are in accordance with the latter discussion. 

	\begin{figure}[h!]
		\begin{center}
		\setlength{\unitlength}{1pt}
			\includegraphics[width = 14 cm]{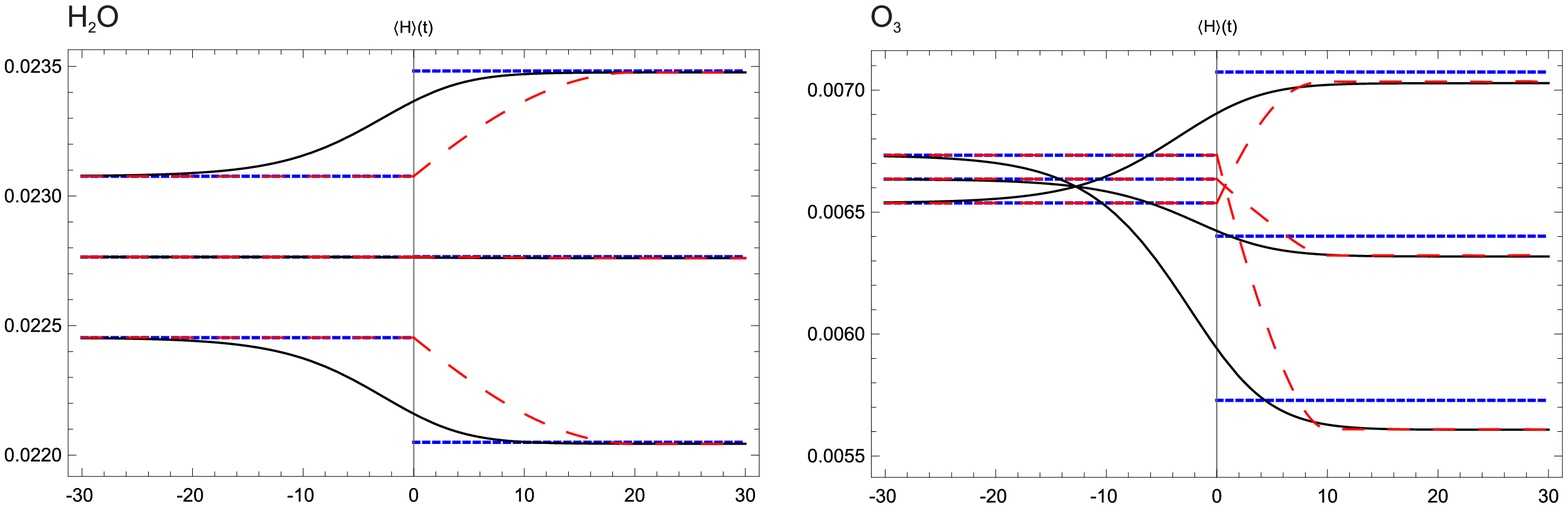}
			\caption{States of the fourth polyad  for molecules with local behaviour. The continuous line stands for the adiabatic change, the short dashed line for the sudden change and the long dashed for the linear change.} 
			\label{Fig-3}
		\end{center}
	\end{figure}

Figure~\ref{Fig-3} is devoted to show the behaviour of the energy levels belonging to a specific polyad, {\it e.g.}, the fourth for symmetric states, during the three different excitation paths.
It is clear that there is a dependence in the final state with the way of excitation. In both molecules the sudden change leads to states at  different energy in comparison to the others excitation types. We believe this behaviour is due to the  approximation taken for the sudden change. 


	\begin{figure}[ h ! ]
		\begin{center}
		\setlength{\unitlength}{1pt}
			\includegraphics[width =13.5 cm]{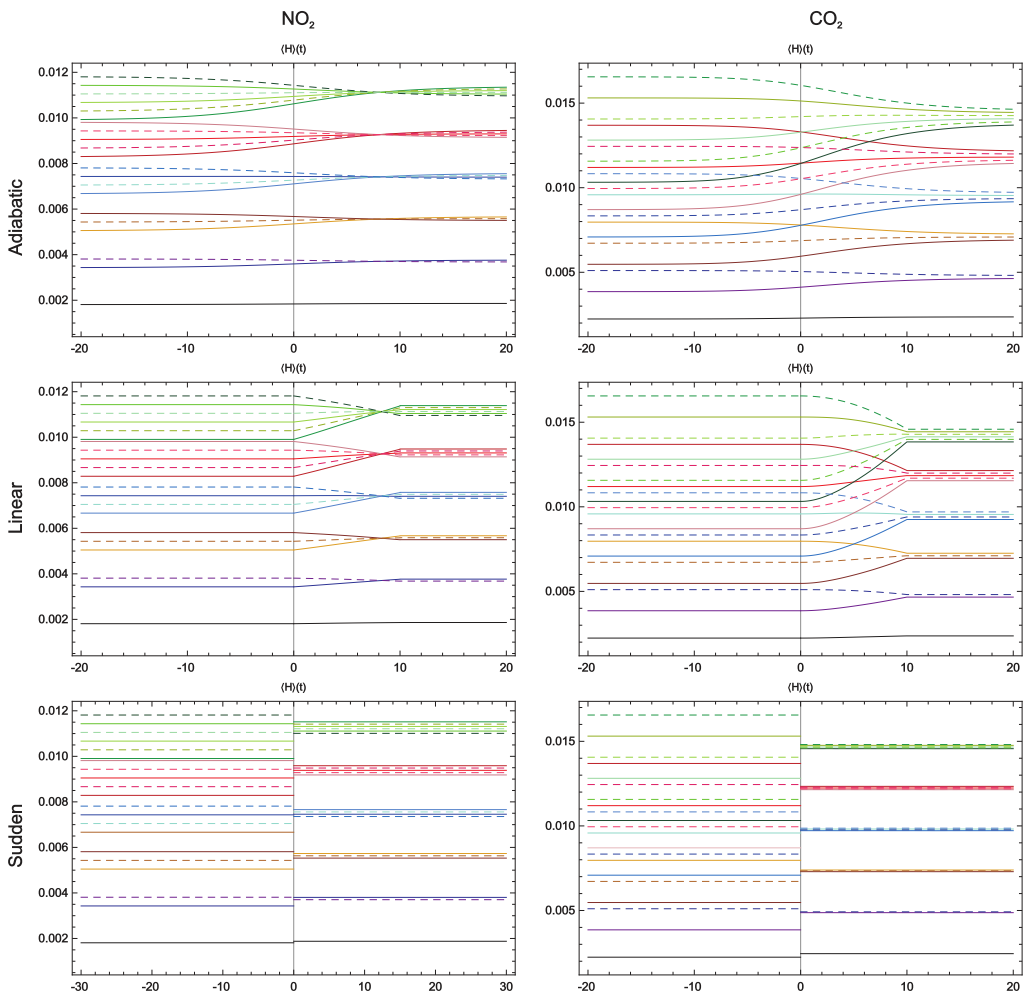}
			\caption{Average  value ($uma\AA^2 /fs$) of Hamiltonian along the time ($fs$) for molecules with normal behaviour. The continuous lines stand for the symmetrical states and dashed lines for the asymmetrical.} 
			\label{Fig-4}
		\end{center}
	\end{figure}

\subsection{Normal-to-Local mode transition}
	
 To illustrate the normal-to-local transition we consider the  molecules NO$_2$ and CO$_2$, where in both cases the angle between bonds  decreases in time until it attains the value $\theta_f = 104.5^\circ$. Following the same procedure described in the above subsection and using the spectroscopic parameters given in  Table~\ref{t1}, the mean values of the Hamiltonian are calculated and displayed in Figure~\ref{Fig-4}. From the right hand side panel of this figure one  appreciates a clear  normal-mode behaviour  at initial time $t_0$ for the CO$_2$ molecule. A local polyad cannot be identified at any energy range. The left panels correspond to the evolution of the energy pattern of the  NO$_2$ molecule, where in  the low lying energy region a local polyad trend can be seen, but immediately the local polyads start approaching making difficult the polyad identification of the states.

In both cases  the excitation induces a local polyad preserving levels trend,  but for the states of the NO$_2$ molecule a collapse into a single state characterized by a local polyad appears due to the same reason explained above.
We can observe  an initial strong normal pattern, followed by a normal-to-local mode transition manifested by groups of local polyads.  Now the $\hat{\mathscr{P}}_{\tiny\mbox{L}}$ can be use to label the states. 

	\begin{figure}[ h ! ]
		\begin{center}
		\setlength{\unitlength}{1pt}
			\includegraphics[width = 8 cm]{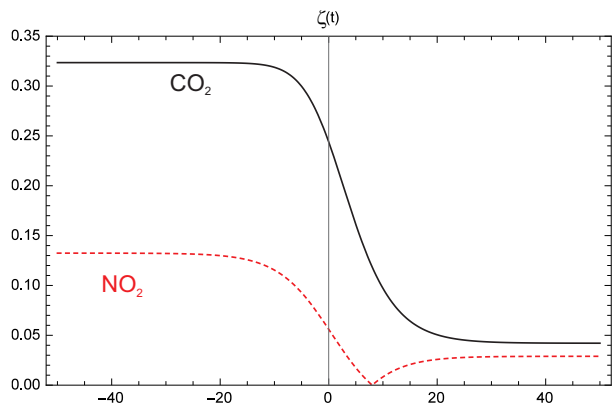}
			\caption{$\zeta(t)$ function for local molecules. Only it is shown the adiabatic change.} 
			\label{Fig-5}
		\end{center}
	\end{figure}

The function $\zeta(t)$ for these molecules is displayed in Figure~\ref{Fig-5}. We  observe that the value of this parameter decreases to $\zeta(t_f)=0.042$ for the CO$_2$ and $\zeta(t_f)=0.023$ for NO$_2$, indicating the local character. These molecules represent a situation where the normal-to-local transition allows a correlation diagram to be established, a feature that makes evident the polyad breaking phenomenon.

	\begin{figure}[ h ! ]
		\begin{center}
		\setlength{\unitlength}{1pt}
			\includegraphics[width = 13 cm]{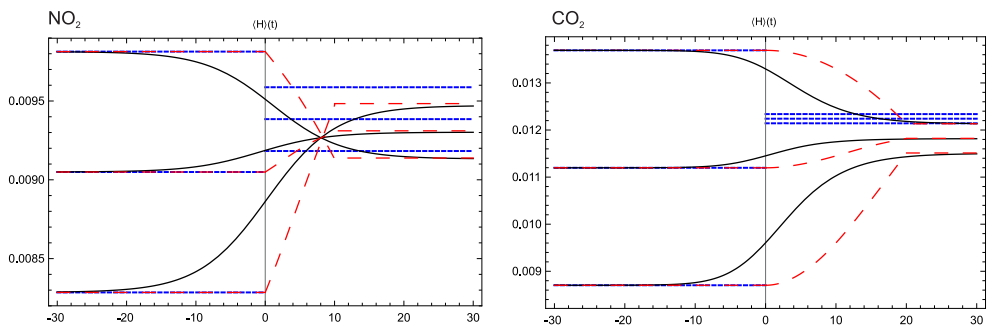}
			\caption{States of the fourth polyad for molecules with normal behaviour. The continuous line stands for the adiabatic change, the short dashed line for the sudden change while the long dashed for the linear change.} 
			\label{Fig-6}
		\end{center}
	\end{figure}

Figure~\ref{Fig-6} shows  the comparison between the three different excitation paths for the three symmetric states belonging to  the  fourth polyad. Again, like in the local case there is a final state dependence of the type of excitation. The sudden change shows a clear different behaviour due to the crude approximation involved.

	\begin{figure}[ h ! ]
		\begin{center}
		\setlength{\unitlength}{1pt}
			\includegraphics[width = 13 cm]{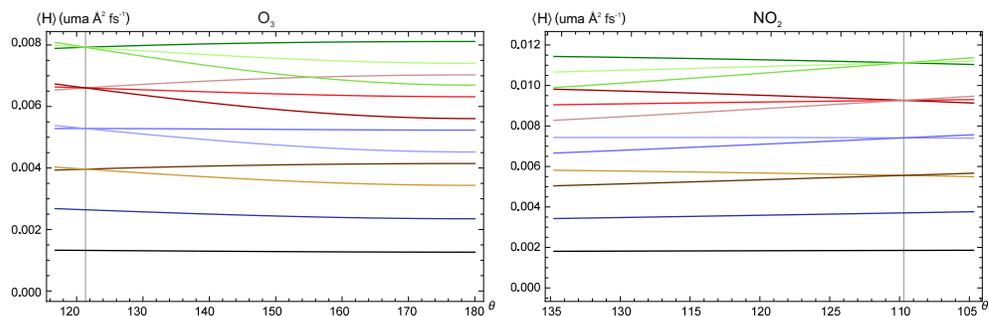}
			\caption{Energy correlation of NO$_2$ molecule in function of $\theta$.} 
			\label{Fig-7}
		\end{center}
	\end{figure}
	
For sake of comparison, in Figure~\ref{Fig-7} the energy correlation for NO$_2$ and $O_3$ molecules as a function of the angle is displayed. These should be compared to the first right panel of Figure~\ref{Fig-1} and to the first left panel of Figure~\ref{Fig-4}, respectively, where  energy levels convergence also appears  due to cancellation effects in $(x_g+x_f)$. The angles corresponding to high degeneracy coincide with the associated time dependent angle.

	\begin{figure}[ h ! ]
		\begin{center}
		\setlength{\unitlength}{1pt}
			\includegraphics[width =13 cm]{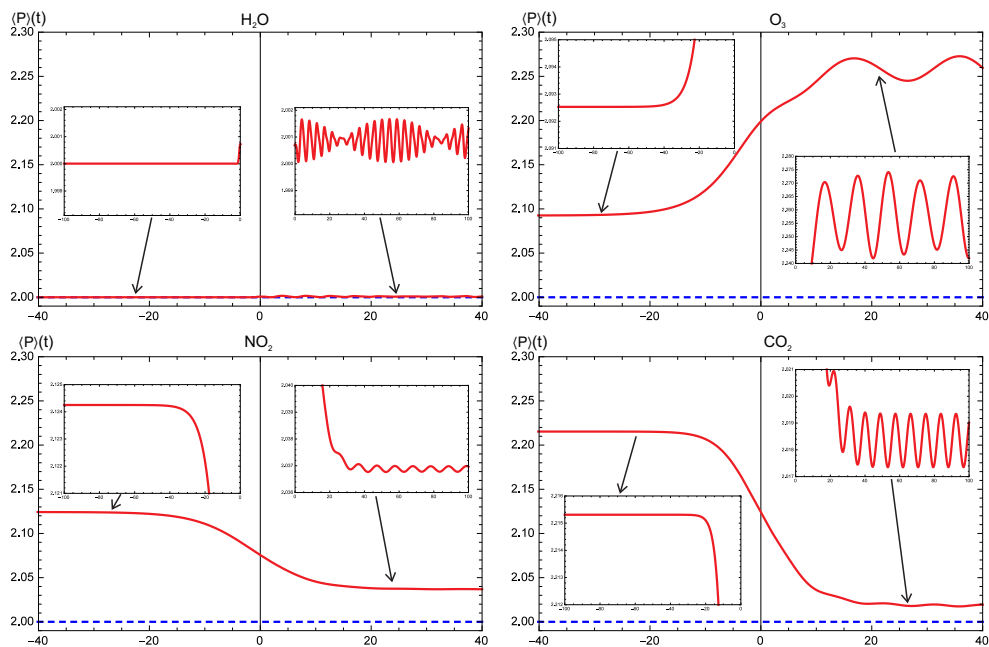}
			\caption{ Time-evolution of the expectation values of the local polyad $\langle \hat{P}_{\tiny \mbox{L}} \rangle(t)$ for the 			molecules under consideration. The expectation value of the normal polyad $\langle \hat{\mathscr P}_{\tiny \mbox{N}} \rangle(t)=2$ is constant. This evolution is only given for the adiabatic change.} 
			\label{Fig-8}
		\end{center}
	\end{figure}

Finally, in Figure~\ref{Fig-8}  
the time evolution of the expectation value of the local polyad $\langle \hat{P}_{\tiny \mbox{L}}\rangle(t)$, given in expression \eqref{Local-Polyad}, is compared to the expectation value $\langle \hat{\mathscr{P}}_{\tiny \mbox{N}} \rangle(t)$. With this analysis we stress the  non conservation in time  of the expectation value of the local polyad $P_{\tiny \mbox{L}}$. In every molecular system $\langle \hat{\mathscr{P}}_{\tiny \mbox{N}} \rangle(t)=2$ since we are considering polyad $2$, but of course this is not the case for $\langle \hat{P}_{\tiny \mbox{L}} \rangle(t)$.
The strongest local character is represented by the water molecule (upper left panel), whose time evolution of the expectation values of the local and normal polyads looks similar. However when the molecule moves to the linear configuration, small   oscillation appears as a consequence of the weak local-to-normal transition. In ozone molecule, although presenting local character, a significant difference between the  $\langle \hat{\mathscr{P}}_{\tiny \mbox{N}} \rangle(t)$ and $\langle \hat{P}_{\tiny \mbox{L}} \rangle(t)$ is present. As the transition to normal character is carried out the change in $\langle \hat{P}_{\tiny \mbox{L}} \rangle(t)$ becomes significant since in this case  $|x_g(t)|$ becomes large. As noticed the oscillation amplitude is significantly larger than in water. Let us now consider the NO$_2$ molecule at the left part of the panel, which presents a larger coupling between the local oscillators . This explains the larger difference between   $\langle \hat{\mathscr{P}}_{\tiny \mbox{N}} \rangle(t)$ and $\langle \hat{P}_{\tiny \mbox{L}} \rangle(t)$.   Since in this molecule the angle is diminished, the normal mode character increases, approaching   $\langle \hat{\mathscr{P}}_{\tiny \mbox{L}} \rangle(t)$ to $\langle \hat P_{\tiny \mbox{N}} \rangle(t)$. Finally the behaviour for the carbon dioxide is displayed in the lower right panel of Figure~\ref{Fig-8}. Here the molecule changes from linear to non linear geometry, manifested by a normal-to-local character. At first because its  strong normal character  the difference between the expected values of the polyads are very different, but as the molecule loses its normal character the expectation values approach. Notice that in all cases oscillations appear, without caring if the transition comes from a normal or local character.

 
\section{Summary and Conclusions}

In this work we have established a transient transition region from local to normal mode behaviour by identifying the condition under which the local polyad $\hat{P}_{\tiny\mbox{L}}$ stop being preserved. This condition is closely related to the suitability of a local model to estimate the force constants at  zeroth order. Hence a polyad defined in a normal mode scheme can be expressed in terms of the polyad in a local mode scheme with additional non polyad  preserving terms. As the splitting between the fundamentals decreases these polyads become equal. This is a remarkable behaviour allowing the polyad breaking process to be studied in different ways. One possibility is to consider the local-to-normal mode transition in parametric way by introducing different tools like probability densities, fidelity, entropy and Poincar\'e sections \cite{marisol}. Another point of view consists in considering a time dependent evolution in the parameters leading to the transition. 

In this contribution we have payed attention to the transition via a time dependent evolution. We have first presented the time dependent solution of two independent harmonic oscillators through the invariants theory. Then a local point of view is analyzed identifying the local time dependent operators associated with the local polyad. In this way the time-dependent local-to-normal mode transition has been established through time dependent polyad breaking phenomenon. 

Several systems has been analyzed, two (H$_2$O and O$_3$) belonging to a local mode behaviour and other two (CO$_2$ and NO$_2$) presenting normal mode behaviour. In all cases the transition has been carried out via sudden, linear and adiabatic ways. The time dependent evolution was introduced through a  change in geometry by modifying the angle in accordance with the time dependence. In our analysis a clear polyad breaking process is manifested in ozone, while in water even a change of geometry does not induced the polyad breaking phenomenon, which is explained by the mass ratios. In  CO$_2$ and NO$_2$ a transient process is manifested to  local  energy grouping by local polyads. Except for the sudden approximation in ozone and NO$_2$ a convergent points of energy level degeneracy by polyad is presented due to the cancellation of the term $(x_f+x_g)$. It has been introduced the locality degree (\ref{zetat}), which provides a   polyad breaking measure from the local point of view.

We believe that this is a non trivial time dependent analysis focusing on the local polyad breaking allowing us to simulate a dynamical study of systems that change geometry during an electronic transition.

\section*{Acknowledges}

This work was supported by DGAPA-UNAM under project IN109113 and CONACyT with reference number 238494. Second author is also grateful for the scholarship (Posgrado en Ciencias Qu\'imicas) provided by CONACyT, M\'exico. We thanks {\it Octavio Casta\~nos Garza } for his comments and the critical  reading of the manuscript.

\appendix

\section{Relation between the local operators} 
\label{Appendix}

In general the local bosonic operators $\{\hat{a}_j, \hat{a}^\dagger_j \}$, $j=1,\,2$, are related to the temporal invariants $\{ \hat{A}_\gamma(t), \hat{A}^\dagger_\gamma(t) \}$, $\gamma = g,\,u$, by means of the matrix relation
{\small
\begin{equation}
\label{A1}
\left(
\begin{array}{c}
  \hat{a}_1   \\
  \hat{a}^\dagger_1  \\
  \hat{a}_2 \\
  \hat{a}^\dagger_2 
\end{array}
\right)
	=\sqrt{\frac{\mu\omega}{2}}
	\left(
	\begin{array}{cccc}
	\chi_g(t) + \frac{i \zeta_g(t)}{\mu\omega} & \chi^\ast_g(t) + \frac{i \zeta^\ast_g(t)}{\mu\omega}  &  \chi_u(t) 					+ \frac{i \zeta_u(t)}{\mu\omega} &  \chi^\ast_u(t) + 	\frac{i \zeta^\ast_u(t)}{\mu\omega}  \\
	
	\chi_g(t) - \frac{i \zeta_g(t)}{\mu\omega} &  \chi^\ast_g(t) - \frac{i \zeta^\ast_g(t)}	{\mu\omega}  &  \chi_u(t) 					- \frac{i \zeta_u(t)}{\mu\omega} &  \chi^\ast_u(t) - \frac{i \zeta^\ast_u(t)}{\mu\omega}  \\
	
	\chi_g(t) + \frac{i \zeta_g(t)}{\mu\omega} &  \chi^\ast_g(t) + \frac{i \zeta^\ast_g(t)}{\mu\omega}  & - \chi_u(t) 				- \frac{i \zeta_u(t)}{\mu\omega} & - \chi^\ast_u(t) - 	\frac{i \zeta^\ast_u(t)}{\mu\omega}  \\
	
	\chi_g(t) - \frac{i \zeta_g(t)}{\mu\omega} &  \chi^\ast_g(t) - \frac{i \zeta^\ast_g(t)}{\mu\omega}  & -\chi_g(t) 					+ \frac{i \zeta_g(t)}{\mu\omega} & - \chi^\ast_g(t) + \frac{i \zeta^\ast_g(t)}{\mu\omega}  \\
	\end{array}
	\right)
		\left(
		\begin{array}{c}
		  \hat{A}_g (t)  \\
		  \hat{A}^\dagger_g (t)  \\
		  \hat{A}_u (t) \\
		  \hat{A}^\dagger_u (t) 
		\end{array}
		\right)\, ,
\end{equation}
}
where
	\begin{equation}
	\chi_{\gamma}(t) = \frac{\alpha_\gamma(t)}{2}e^{-i\phi_\gamma}\, , \quad 
	\zeta_\gamma(t) = \frac{e^{-i\phi_\gamma}}{2 i} \left( \frac{1}{\alpha_\gamma(t)} + i\frac{\dot{\alpha}_\gamma}				{G_{\gamma\gamma}} \right) \, .
	\end{equation}
From (\ref{A1})  we are able to find the temporal evolution of the local polyad $\hat{P}_{\tiny\mbox{L}} = \hat{n}_1 + \hat{n}_2$, where $\hat{n}_1 $ and  $\hat{n}_2$ will be functions of the operators $\{ \hat{A}_\gamma(t), \hat{A}^\dagger_\gamma(t) \}$. In addition, it is possible to obtain the average   values associated with the local polyad through the eigenstates of the operators $\{ \hat{A}_\gamma(t), \hat{A}^\dagger_\gamma(t) \}$, i.e.
	\begin{equation}
	\langle \hat{P}_{\tiny\mbox{L}} \rangle = \langle n_g,n_u, t| \hat{P}_{\tiny\mbox{L}} | n_g,n_u,t \rangle \, .
	\end{equation}
The result of this calculation is given by the equation (\ref{Local-Polyad}).

It is also possible to express the set $\{\hat{a}_j, \hat{a}^\dagger_j \}$ in terms of the operators $\{\hat{A}_{j \gamma}, \hat{A}^\dagger_{j\gamma} \}$ through the relation (\ref{ag12}). In particular in the local limit this relations takes the form  (\ref{a12gu}), i.e.
	{\small
\begin{equation}
\left(
\begin{array}{c}
\hat{A}_g (t)  \\
\hat{A}^\dagger_g (t)  \\
\hat{A}_u (t) \\
\hat{A}^\dagger_u (t) 
\end{array}
\right)
	=\frac{1}{\sqrt{2}}
	\left(
	\begin{array}{cccc}
	1 & 0  &  1 &  0  \\
	0 & 1  & 0 & 1  \\
	1 & 0  & -1 & 0  \\
	0 & 1  & 0 &-1  \\
	\end{array}
	\right)
		\left(
		\begin{array}{c}
		  \hat{A}_1 (t)  \\
		  \hat{A}^\dagger_1 (t)  \\
		  \hat{A}_2 (t) \\
		  \hat{A}^\dagger_2 (t) 
		\end{array}
		\right)\, ,
\end{equation}
}
and taking into account that in this limit  $\alpha_g(t) = \alpha_u(t) = \alpha(t)$, then
{\small
\begin{equation}
\left(
\begin{array}{c}
  \hat{a}_1   \\
  \hat{a}^\dagger_1  \\
  \hat{a}_2 \\
  \hat{a}^\dagger_2 
\end{array}
\right)
	=\sqrt{\mu\omega}
	\left(
	\begin{array}{cccc}
	\chi(t) + \frac{i \zeta(t)}{\mu\omega} & \chi^\ast(t) + \frac{i \zeta^\ast(t)}{\mu\omega}  & 0 & 0  \\
	
	\chi(t) - \frac{i \zeta(t)}{\mu\omega} &  \chi^\ast(t) - \frac{i \zeta^\ast(t)}{\mu\omega}  & 0 & 0  \\
	
	0 & 0 & \chi(t) + \frac{i \zeta(t)}{\mu\omega} &  \chi^\ast(t) + \frac{i \zeta^\ast(t)}{\mu\omega}  \\
	
	0 &  0 & \chi(t) - \frac{i \zeta(t)}{\mu\omega} &  \chi^\ast(t) - \frac{i \zeta^\ast(t)}{\mu\omega}  \\
	\end{array}
	\right)
		\left(
		\begin{array}{c}
		  \hat{A}_1 (t)  \\
		  \hat{A}^\dagger_1 (t)  \\
		  \hat{A}_2 (t) \\
		  \hat{A}^\dagger_2 (t) 
		\end{array}
		\right)\, ,
\end{equation}
}
from which we can obtain the temporal evolution of the local polyad 
$\hat{P}_{\tiny\mbox{L}}$ in the local limit. The average values of this oparator may be calculated by means of the eigenstates of  $\{ \hat{A}_j, \hat{A}^\dagger_{j}\}$, which yields
	\begin{equation}
	\langle n_1,n_2,t | \hat{P}_{\tiny\mbox{L}} |n_1,n_2,t \rangle = \frac{1}{2\hbar}\sum_{j=1,2}\left[ \mu\omega\sigma^2_{q_j} + \frac{1}{\mu\omega}		\sigma^2_{p_j} \right] - 1
	\end{equation}
which is precisely  ({\ref{Local-Polyad}) in the local limit.


\vskip 1cm


\end{document}